\documentstyle[12pt]{article}
\begin{document}
\vskip 72pt
\centerline{\bf SUPERHEATED DROP AS A NEUTRON SPECTROMETER} 
\vskip 36pt
\centerline{\bf Mala Das, B. K. Chatterjee, B. Roy\footnote
[1]{Author for communication : Tel.no.-350 2402/03, Ext.-305, Fax no.- 91 033 350 6790, 
e-mail-biva@boseinst.ernet.in} and S. C. Roy}
\vskip 36pt
\centerline{\it Department of Physics, Bose Institute}
\centerline{\it 93 / 1 A. P. C. Road, Calcutta 700009, India}
\vskip 36pt
\begin{abstract}

        Superheated drops are known to vaporise when exposed to
energetic nuclear radiation since the discovery of bubble chamber. The
application of  superheated drops in neutron research specially in
neutron dosimetry is a subject of intense research for quite sometime.
As the degree of superheat 
increases in a given liquid, less and less energetic neutrons are
required to cause nucleation. This property of superheated liquid makes
it possible to use it as a neutron spectrometer. Neutron detection
efficiency of superheated drops made of R12 exposed to Am-Be neutron
source have been measured over a wide range of temperature -17$^o$C to
60$^o$C and the results have been utilized to construct the energy
spectrum of the neutron source. This paper demonstrates that a suitable 
neutron spectrometer may be constructed by using a single liquid and 
varying the temperature of the liquid suitably at a closer grid.
\vskip 36 pt


\vskip 36 pt

{\bf Keywards} : SDD, R12, temperature dependence, neutron, efficiency, 
spectrometry.  

\end{abstract}

\noindent

\pagebreak

\noindent{\bf 1. INTRODUCTION}

\vskip 24 pt

        A fluid kept in the liquid state above its boiling temperature
is called superheated. Application of superheated liquid to detect
ionizing radiation is well known from the times of bubble chamber[1].
The resurgence of its use has been observed from the late seventies[2,3]
and the investigations on the subjects in the last two decades turned
into an almost maturing technology especially to detect neutrons and
more recently to detect photons. The suitability of using superheated
drops as a neutron dosimeter[4,5,6,7,8] has already been established. 
The superheated drops are now commercially available as neutron and photon 
dosimeter in the trade name superheated drop detector (SDD) \footnote
[2]{Superheated Drop Detector is the registered trademark of Apfel
Enterprises Inc, NewHaven, CT, USA} and bubble detector (BD). \footnote
[3]{Bubble Detector is the registered trademark of Bubble Technology
Industries Ltd.} In a superheated liquid minimum energy (threshold)
required to nucleate decreases as the degree of superheat increases.
The degree of superheat of a liquid could be defined simply by the
difference of ambient temperature above the boiling point of the
liquid. Therefore, liquids with lower boiling points possess higher
degree of superheats at a given ambient temperature (above their
boiling points) and as the ambient  temperature increases the given
liquid becomes more and more superheated. This property of the
superheated liquid are being utilised to develop neutron dosimetry and
neutron spectrometry[2,3,9,10]. There are two distinct types of
methodologies used in developing neutron spectrometer. In one, a
collection of superheated samples made of liquids with different
boiling points (i.e. with different threshold neutron energies) are
utilised[11], while in the other, two liquids are chosen and the
temperature of the liquids are varied at four different temperatures to
obtain eight sets of threshold energies[12,13,14] (equivalent to eight
different samples with different boiling points). The temperature
variation method is superior than using samples with different boiling
points. By controlling the temperature of the sample one can, in
principle, change the threshold neutron energy at any desired level 
(equivalent to using 'finer' windows to scan the spectrum),
while in the other method one is limited by the availability of liquids
with lower boiling points (equivalent to using 'coarser' windows). 
In the present work we measured the
detection efficiency of a single sample made of R-12(Dichlorodifluoromethane : 
C${Cl_2}$${F_2}$), which is known to
be sensitive to neutrons of energies from thermal to tens of MeVs, by
(almost) continuously changing temperatures over a wide range. The
response of the sample to Am-Be neutrons have been measured at about
thirty different temperatures in the range which is equivalent to using 
thirty
different samples with thirty different boiling points (the experiment
has been actually performed more than 50 different temperatures in the range
-17$^oC$ to about 60$^oC$ but it has been observed that the sample
started responding from about 0.5$^oC$ to Am-Be neutrons). In addition
to the advantage of using single liquid, controlling the temperature 
enables one to scan the energy spectrum by finer 'windows' thereby 
improving the inherent energy resolution of the spectrometer when compared 
with other such spectrometers using superheated liquid.

\vskip 36 pt

\noindent{\bf 2. PRINCIPLE OF OPERATION}
\vskip 24 pt

        The superheated state of the liquid is a metastable state and
the nucleation in this state can be initiated by the presence of
heterogeneous nucleation sites such as air bubbles, solid impurites,
gas pockets etc. or by radiation interactions. The nucleation in
superheated state starts with the formation of a critical sized vapour
embroy. The free energy required to form a spherical vapour bubble  of
radius r in a liquid is given by  
\begin{equation}
G = 4 {\pi} {r^2}  {\gamma\left(T\right)} - {4\over3} {\pi} {r^3}
{\left({p_v} -{p_o}\right)} 
\end{equation}                   

\noindent where $\gamma(T)$ is the liquid-vapour interfacial tension,
$p_v$ is vapour pressure of the superheated liquid and $p_o$ is the
ambient pressure. The difference $p_v$ -$p_o$ is called the {\it degree
of superheat} of a given liquid. One can see from equation (1) that G
is maximum at 
\begin{equation} 
r = 2 {\gamma\left(T\right)} / {\left({p_v} -{p_o}\right)} = {r_c}     
\end{equation}

\noindent where $r_c$ is called the {\it critical radius}. When a
bubble grows to  the size of the critical radius it becomes
thermodynamically unstable and grows very fast till the entire liquid
droplet vaporises. 

     The minimum amount of energy (W) needed to form a vapour bubble of
critical size $r_c$ as given by Gibbs[15] from reversible
thermodynamics is 
\begin{equation}
W = 16 {\pi} {\gamma{^3}\left(T\right)}/3
{{\left({p_v}-{p_o}\right)}^2}
\end{equation}           

\noindent where ${\gamma\left(T\right)}$ = C ${\left({T_c} - T - d
\right)}$  with ${T_c}$ the critical temperature of the liquid and 
C and d are constants [16]. 

With increase in temperature, since the degree of superheat
($p_v$-$p_o$) increases and $\gamma(T)$ decreases, the minimum energy
($W$) required for vapour bubble nucleation will be less. The variation
of $W$ with temperature for superheated drops of R-12 is shown in
figure 1. Therefore $W$, the threshold energy for nucleation depends on
the type and the temperature of the liquid. When a neutron of energy
{$E_n$} interacts with a nucleus of atomic weight A, the maximum energy
that can be transferred to the nucleus from the neutron is through the
elastic head on collision and is given by,
\begin{equation}
{E_i} = 4 A {E_n}/{{\left(A+1\right)}^2}        
\end{equation}
        
        After receiving the energy, the nucleus is scattered from its
atom and moves through the liquid losing its energy through Coulombic
interaction until it comes to rest. For a given neutron energy,
different nuclei of the liquid will receive different amount of energy,
depending on their atomic weight. The ion with the highest value of
linear energy transfer (LET) or ($dE/dx$) in the liquid, will
play the major role in vapour nucleation[17]. The energy deposited
along that part of the ion's path (L) corresponding to about twice the
critical radius contributes significantly to bubble formation[17,18].
For nucleation to occur this deposited must exceed $W$ the minimum
energy required for bubble formation.

Usually most of the energy is lost into heat and a very small fraction
of the deposited energy is utilised in nucleation and W/${E_c}$ is
called the thermodynamic efficiency (${{\eta}_T}$ of nucleation [17]. 

\begin{equation}
                W = k {r_c} dE/dx = 2{{\eta}_T}{r_c} dE/dx
\end{equation}

where $k$ (constant) equals twice the thermodynamic efficiency ($\eta$).
Hence, in the equation
\begin{equation}
W/r_c = k dE/dx
\end{equation}
\noindent relates the threshold energy (corresponding to the $dE/dx$)
for nucleation to the ambient temperature (corresponding to $W/{r_c}$).
This enables us to convert the temperature scale of superheated drops
to the (threshold) energy scale of incident neutrons. Therefore by
varying the ambient temperature of the superheated drops, one can
observe the variation of the SDD response at different neutron energies
which has been used in neutron spectrometry. 

\vskip 36 pt

\noindent{\bf3. EXPERIMENT}
\vskip 24 pt

        The experiment was performed with superheated drops of R12 at
different temperatures by using the volumetric method, described
by Das {\it et al.}[19]. The vial containing the sample was
connected to a graduated horizontal glass tube with a small coloured
water column as marker. The vapourization of a liquid drop displaces
the water column by the distance corresponding to the volume of vapour
formed. The details of the preparation of the sample is given elsewhere[20]. 

        If neutrons  of flux $\psi$ are incident on superheated drops
of volume  $V$,  liquid of density$\rho_L$ and molecular weight $M$ the
vaporization rate is given by 

\begin{equation}
{\frac{dV}{dt}} = V\psi {\frac{N_A\rho_L}{M}} \eta d \sum{n_i\sigma_i}
\end{equation}

\begin{tabbing}
\noindent where \= N$_A$ \= = \= Avogadro Number\\
        \> d \> = \> average droplet volume\\
        \> n$_i$ \> = \> weight factor of the ith element in the
molecule whose\\
         \> \>   \> neutron nucleus elastic scattering cross section is
$\sigma_i$\\
        \> $\eta$ \> = \> efficiency of neutron detection.
\end{tabbing}   

        Due to nucleation by neutrons, the displacement of the water
column along a horizontal glass tube was measured as a function of
time. The procedure of calculating $\eta$ from the measured
displacement of water column has been explained in detail in one of our
recent publications[19].    
   
        The temperature of the sample was controlled by an
indigeneously made temperature controller. For low temperature
measurements the sample was placed in an alcohol bath sitting on the
top of a cold finger dipped in liquid nitrogen. The upper part of the
finger was wrapped with heating tape and by applying different voltages
to the tape, different steady temperatures of the bath can be achieved.
For measurements of higher temperatures the same setup was used without
the liquid nitrogen. The fluctuation of temperature in these
measurements was found to be within $\pm 0.1^oC$.       

         In the experiment the ambient temperature of the sample was
increased slowly from low to any desired higher value and the
nucleation rate was measured at each temperature. The measurement was 
performed from in a temperature range of -17$^oC$ to about 60$^oC$ in a
close grid necessary to obtain the energy spectrum of the source from
temperature. The nucleation due to background radiation and due to
other fluctuations has been subtracted. The liquid was observed to
become unstable due to spontaneous nucleatation at about 60$^oC$. 
        
\vskip 36 pt

\noindent{\bf 4. EXPERIMENTAL RESULTS AND COMPUTATION}
{OF THE NEUTRON ENERGY} 
\vskip 24 pt
        
        The variation of neutron detection efficiency ({$\eta$}) for
R12 with temperature in presence of neutrons from Am-Be neutron source
is presented in figure 2. The solid line in figure 2 is the spline smoothing 
of the efficiency data at different temperatures. The uncertainties presented
in the figure are the total experimental uncertainties of estimating $\eta$. 
The derivative of efficiency, {d$\eta$/dT}
against temperature is shown in figure 3. Now one has to estimate the
equivalence of the energy of the detected neutrons with the temperature
of the SDD sample. One way to do it is to expose the sample at
different temperatures to different monoenergetic neutron sources and
to note the threshold neutron energies for nucleation [12,14]. A novel
approach has been used in this work.

        As has been presented in Section 2, the different nuclei of the
superheated liquid would receive different amount of energy and they
must have different dE/dx. In case of R-12 containing C, Cl and F,
dE/dx of these ions with different neutron energy are presented in
Figure 4. From Figure 4, it is clear that the dE/dx values of C, Cl and
F are comparable in the neutron energy of our interest and we take the
average value of dE/dx of all the ions using the equation below

\begin{equation}
{\left(dE/dx\right)}_{average}
=\sum{{n_i\sigma_i}{\left(dE/dx\right)_i}}/\sum{n_i\sigma_i} 
\end{equation}

\noindent where ${n_i}$ is the number of ions of the i-th element in a
R12 molecule, {$\sigma_i$}  is the neutron-nucleus elastic scattering
cross section and {$(dE/dx)_i$} is the LET of the i-th ion. The
variation of average dE/dx as a function of neutron energy is shown in
figure 5.

        From equations (3) and (5), we obtain
\begin{equation}
{\left(dE/dx\right)} = 8 {\pi}
{\gamma{^2}\left(T\right)}/3k{\left({p_v} -{p_o}\right)}             
\end{equation}

\noindent From the equation above the dE/dx has been plotted against
temperature for different arbitrary values of $k$ of which only four
such plots ($k$=1, 0.1, 0.05, 0.0195) are presented in figure 6 using
the equivalence between the temperature of the sample and the incident
neutron energy from figure 5. The variation of threshold neutron energy
for nucleation with temperature of the sample for different $k$ has
been studied of which four such variations are shown in figure 7. With
the optimum value of $k$ the temperature axis of the figure 3 has been
converted to neutron energy and the resulting spectrum was fitted with
the peak neutron energy of the $^{241}$Am-Be neutron spectrum.
The best fit is obtained for k equals 0.0195. The analysis has been 
performed upto a maximum temperature of 42.5$^oC$. The final neutron energy
spectrum of $^{241}$Am-Be source obtained from our analysis is shown in
figure 8.   
        
        If L = 2{$r_c$} is taken as the distance in the ion's path
which contributes  significantly in nucleation of superheated drops
[17,18] in calculating the thermodunamic efficeincy of nucleation, our
experimental analysis produces the value close to 0.01.  It may be
noted here that Apfel {\it et al.}[17] obtained this value ranging from
0.03 to 0.05.

\pagebreak
\noindent{\bf 5. DISCUSSION}
\vskip 24 pt

        The result shows that {$\eta$} increases with temperature. At
low temperature, the threshold energy for nucleation (W) is high, which
indicates that a larger energy is required to cause nucleation.
According to equation 3 as temperature increases, W decreases and more
and more neutrons from the low energy range of the spectrum are taking
part in nucleation. So {$\eta$} increases with temperature. The sharp
increase of {$\eta$} near 25$^o$C corresponds to detection of neutrons
with energies ranging from highest available to those at the peak of
the spectrum. At high temperature when all the neutrons of the spectrum
contribute in nucleation, {$\eta$} should be constant with temperature.
But at about 45$^o$C, {$\eta$} increases again. We suspected that the 
sample becomes sensitive to gamma rays coming out of the Am-Be source.
In a separate experiment, we indeed observed that R-12 becomes
sensitive to gamma rays at about 45$^oC$. Since, as mentioned before, the
analysis has been performed upto a maximum temperature of 42.5$^oC$, the 
contribution due to gamma rays is absent in the measured energy spectrum 
in this work.  
	
	 Figure 3 shows that the d${\eta}$/dT vs. T graph resembles the
neutron energy spectrum of $^{241}$Am-Be where the second peak
corresponds to the gamma sensitivity of the sample. The ambient
temperature of the superheated drops was converted to the energy of the
neutrons following the method described in Section 4. So by using
superheated drops at different temperatures, it is possible to obtain
the neutron energy spectrum. This indicates important use of SDD in
neutron spectrometry. The maximum uncertianty in neutron energy as could 
be found from figure 7 is within 5${\%}$ in the entire region of our 
investigation. This method can be used to determine any other
neutron energy outside the present range, only then one has to consider
the ion with maximum dE/dx and the rest of the analysis 
is same as this. The present study also helps to select the suitable
material (liquid) for a given neutron energy spectrum.
        
        It has been observed in this experiment that the nucleation
rate of superheated drops rapidly changes for samples exposed to {\it
thermal shock} compared to samples whose temperature was changed
slowly. The liquid appeared to be more fragile when temperature was
changed rapidly. Though this is not quite unexpected in the exact
physics of this phenomenon, why the liquid becomes more fragile under
heat shock, requires further investigation.     

\pagebreak
        
\noindent{\bf REFERENCES}

\vskip 24 pt

\noindent{1. D.A. Glaser, {\it Phys. Rev. A} {\bf 87}, 665 (1952).}\\
\noindent{2. R.E. Apfel, {\it US Patent} 4,143,274 (1979).}\\
\noindent{3. R.E. Apfel, {\it Nucl. Inst. Meth.} {\bf 162}, 603(1979).}\\
\noindent{4. R.E. Apfel and S.C. Roy, {\it Nucl. Inst. Meth.} {\bf 219}, 582
(1984).}\\
\noindent{5. R.E. Apfel and Y.C. Lo, {\it Health Phys.} {\bf 56}, 79 (1989).}\\
\noindent{6. R.E. Apfel, {\it Rad. Prot. Dos. } {\bf 44}, 343 (1992).}\\
\noindent{7. S.C. Roy, R.E. Apfel and Y.C. Lo, {\it Nucl. Inst. Meth.} {\bf
A255}, 199 (1987).}\\
\noindent{8. H.Ing, {\it Nuclear Tracks.} {\bf 12}, 49 (1986).}\\
\noindent{9. R.E. Apfel, {\it Nucl. Inst. Meth.} {\bf 179}, 615 (1981).}\\
\noindent{10. K. Chakraborty, P. Roy, S.G. Vaijapurkar and S.C. Roy, {\it Proc.
of 7th National Conference on Particles and Tracks}, Jodhpur pp 133
(1990).}\\
\noindent{11. H. Ing, R. A. Noulty and T. D. Mclean, {\it Rad. Meas.}{\bf 27}, 1
(1995).}\\
\noindent{12. F. d'Errico, W. G. Alberts, G. Curzio, S. Guldbakke, H. Kluge, and
M. Matzke, {\it Rad. Proc. Dos.} {\bf 61}, 159 (1995).}\\ 
\noindent{13. F. d'Errico, R. E. Apfel, G. Curzio, E. Dietz, G. F.Gualdrini, S.
Guldbakke, R. Nath, B. R. L. Siebert, {\it Rad. Proc. Dos.} {\bf70}, 1
(1997).}\\
\noindent{14. F. d'Errico, W. G. Alberts and M. Matzke, {\it Rad. Proc.Dos.}
{\bf 70}, 103 (1997).}\\  
\noindent{15. J. W. Gibbs, Translations  of  the  Connecticut  Academy III, 
p.108 (1875).}\\
\noindent{16. F. H. Newman and V. H. L. Searle, The general properties of 
matter(fifth ed.), p.189 (1985).}\\ 
\noindent{17. R. E. Apfel, S. C. Roy and Y. C. Lo, {\it Phys. Rev. A}{\bf 31},
3194 (1985).}\\ 
\noindent{18. M. J. Harper and M.E. Nelson, {\it Radiat. Prot. Dosim.} {\bf 47
}, 535 (1990).}\\
\noindent{19. Mala Das,B. Roy, B. K. Chatterjee and S. C. Roy, {\it Rad.Meas.}
{\bf 30}, 35 (1999).}\\
\noindent{20. B. Roy, B. K. Chatterjee and S. C. Roy, {\it Rad. Meas.} {\bf 29},
173 (1998).}\\
 
\pagebreak

\noindent{\bf FIGURE CAPTIONS}

\vskip 24 pt

\noindent{\bf Fig. 1:} Variation of threshold energy (W) required for
nucleation in R12 as a function of temperature (T). 
\vskip 36 pt

\noindent{\bf Fig. 2:} Observed variation of neutron detection
efficiency ({$\eta$}) as a function of temperature (T) in R12.
\vskip 36 pt

\noindent{\bf Fig. 3:} Variation of the derivative of neutron detection
efficiency (d{$\eta$}/dT) as a function of temperature (T).
\vskip 36 pt

\noindent{\bf Fig. 4:} Variation of stopping power (dE/dx) of different
ions (C, Cl, F) in Freon-12 as a function of neutron energy.
\vskip 36 pt

\noindent{\bf Fig. 5:} Variation of average stopping power
${(dE/dx)_{average}}$ over three different ions in Freon-12 as a
function of neutron energy. 

\vskip 36 pt

\noindent{\bf Fig. 6:} Variation of stopping power (dE/dx) of ion in
R12 as a function of temperature (T) of the sample, for different
arbitrary values of $k$. 
\vskip 36 pt

\noindent{\bf Fig. 7:} Variation of neutron energy as a function of
temperature (T) of the sample, for different $k$.
\vskip 36 pt

\noindent{\bf Fig. 8:} The neutron energy spectrum of $^{241}$Am-Be
obtained from the experiment.

\end{document}